\documentclass[aps,prl,twocolumn,showpacs,floatfix,preprintnumbers,amsmath,amssymb,superscriptaddress,reprint]{revtex4-1}

\usepackage[colorlinks,bookmarks=false,citecolor=blue,linkcolor=blue,urlcolor=blue]{hyperref}
\usepackage[all]{hypcap}   

\usepackage{amsmath,amssymb}
\usepackage{graphicx}
\graphicspath{{figures/}{/}}
\usepackage{dcolumn}
\usepackage{bm}
\usepackage[capitalise,compress]{cleveref}
\crefname{section}{Sec.}{Secs.}
\Crefname{section}{Section}{Sections}

\usepackage{verbatim}
\usepackage{color}
\usepackage{soul}
\usepackage{placeins}  
\usepackage{flafter}     

\usepackage{dsfont}

\usepackage{color}

\usepackage[final]{pdfpages}
\usepackage{pgffor}
\makeatletter
\AtBeginDocument{\let\LS@rot\@undefined}
\makeatother

\newcommand{\HIDDEN}[1]{}

\makeatletter
\let\Hy@backout\@gobble
\makeatother

\begin{document}

	\title{Hidden Quantum Criticality and Entanglement in Quench Dynamics}
	
	\author{Sanku Paul}
	\affiliation{Department of Physics and Astronomy, Michigan State University, East Lansing, Michigan 48824, USA}
	\affiliation{Kavli Institute for Theoretical Physics, University of California, Santa Barbara, CA 93106-4030, USA}
	
	\author{Paraj Titum}
	\affiliation{Johns Hopkins University Applied Physics Laboratory, Laurel, Maryland 20723, USA}
	\affiliation{Kavli Institute for Theoretical Physics, University of California, Santa Barbara, CA 93106-4030, USA}
	\affiliation{Joint Quantum Institute, NIST/University of Maryland, College Park, Maryland 20742, USA}
	
	\author{Mohammad F. Maghrebi}
	\affiliation{Department of Physics and Astronomy, Michigan State University, East Lansing, Michigan 48824, USA}
	\affiliation{Kavli Institute for Theoretical Physics, University of California, Santa Barbara, CA 93106-4030, USA}
	
	
	\begin{abstract}
		Entanglement exhibits universal behavior near the ground-state critical point where correlations are long-ranged and the thermodynamic entropy is vanishing. 
		On the other hand, a quantum quench imparts extensive energy and results in a build-up of entropy, hence no critical behavior is expected at long times. In this work, we present a new paradigm in the quench dynamics 
		of integrable spin chains which
		exhibit a ground-state order-disorder phase transition at a critical line. Specifically, we consider a quench along the critical line which displays a volume-law behavior of the entropy and exponentially decaying correlations; however, we show that quantum criticality is hidden in higher-order correlations and becomes manifest via measures such as the mutual information and logarithmic negativity.  
		Furthermore, we showcase
		the scale-invariance of the R\'{e}nyi mutual information between disjoint regions as further evidence for genuine critical behavior. We attribute the emerging universality to the vanishing effective temperature of the soft mode in spite of the quench. Our results are amenable to an experimental realization on different quantum simulator platforms, particularly the Rydberg simulators.
	\end{abstract}
	
	\maketitle

	
	Entanglement
	characterizes non-classical correlations in a quantum state and has a wide variety of applications in quantum computing, networking, and metrology.  
	Furthermore, entanglement provides a powerful diagnostic for quantum phase transitions~\cite{Osborne02, Osterloh02, Vidal_2003,Vidal_2004}.
	For a pure state, it can be quantified 
	by the von-Neumann entropy $S_A$ of a given subsystem $A$.
	For a one-dimensional (1D) spin chain, we generically have~\cite{Calabrese_2004}
	\begin{equation}
	\label{Svn_exp}
	S_A=a |A| +b \ln |A|+ \rm{const},
	\end{equation}
	with $|A|$ the subsystem size and $a$, $b$ constants independent of $|A|$.
	Highly excited states (or finite-temperature states) typically obey a \emph{volume law} with $a\neq 0$, reflecting the thermodynamic entropy of the state. In contrast,  the ground state of gapped Hamiltonians exhibit an \emph{area law} where 
	$S_A$ is a constant independent of system size (i.e., $a$, $b$=0).
	In both cases, only short-range correlations are present in the state.
	On the other hand, a leading logarithmic term emerges at a quantum critical point in the ground state with $b$ a universal coefficient (while $a=0$). For a conformal field theory~(CFT), this coefficient is $b=c/3$ with $c$ the central charge~\cite{Calabrese_2009_conformal}.
	The logarithmic  
	scaling of entanglement entropy is a powerful indicator of criticality which is more conventionally diagnosed with power-law correlations~\cite{zinn2021quantum}. 
	Conversely, the absence of universal logarithms means no critical behavior.
	For instance, thermal states do not exhibit logarithmic corrections~\cite{Wolf_2008}, consistent with the fact that there are no 1D phase transitions at finite temperature~\cite{landau2013course}.
	
	How does this paradigm change for non-equilibrium states? Here, we consider the stationary states of an isolated system upon a sudden quench. 
	Generic quantum systems are widely believed to thermalize~\cite{DAlessio2016,Gogolin2016,Srednicki1999-eth}, hence the volume law emerges while universal logarithms do not~\cite{Marko_2008,Lauchli_2008,Schachenmayer2013,Kim2013}. In contrast, integrable systems evade thermalization and approach a stationary state at long times~\cite{Rigol2011,Konik2012,Fagotti2013,Sirker2014,Vidmar_2016}. But even in this case, the long-time stationary states typically exhibit exponentially decaying correlations~\cite{Calabrese_2012_I,Calabrese_2012_II} and extensive energy/entropy~\cite{Calabrese_2005,Calabrese_2007,Chiara_2006,Alba7947,Alba_2019,Vidmar_2018}. A rather special exception is coupled harmonic oscillators where subleading logarithms appear in the dynamics ~\cite{Nezhadhaghighi_2014,Cotler2016}, due to their zero modes harboring an arbitrary large entropy~\cite{Cotler2016,Unruh1990}. Current-carrying steady states coupled to two different baths could also lead to subleading logarithms~\cite{Eisler14, Ribeiro17}. 
	However, generic settings of quench dynamics do not exhibit criticality at late times, akin to thermal states, hence $a\ne0$ while $b=0$.

	In this work, we present a new paradigm for entanglement 
	and criticality in the long-time stationary state of quench dynamics. 
	We consider the anisotropic XY chain as a paradigmatic integrable model and show that a quench along the critical line leads to a volume law plus logarithmic corrections. While the latter indicate criticality, characteristic correlation functions decay exponentially. 
	However, we show that a form of \textit{quantum} criticality is \textit{hidden} in higher-order correlation functions, captured via quantum information measures such as mutual information and logarithmic negativity.
	We attribute this critical behavior to the vanishing effective temperature of the soft mode at criticality.
	Our work uncovers a new paradigm of quantum phase transitions displayed by quantum information measures, but invisible to correlation functions and local observables.

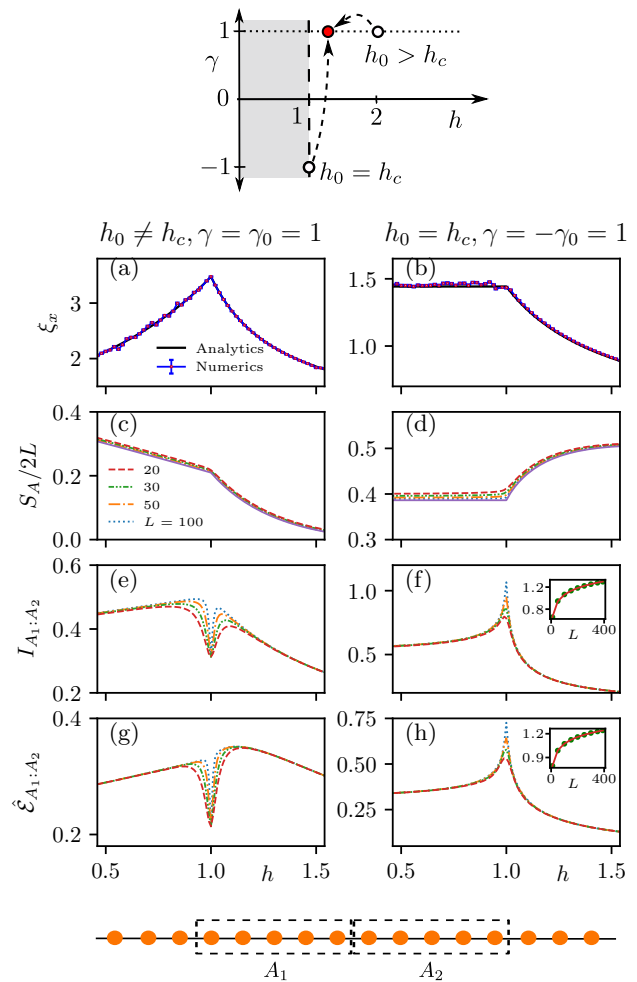
\begin{figure}
		\hspace{0.5cm}
		\includegraphics[width=0.442\linewidth]{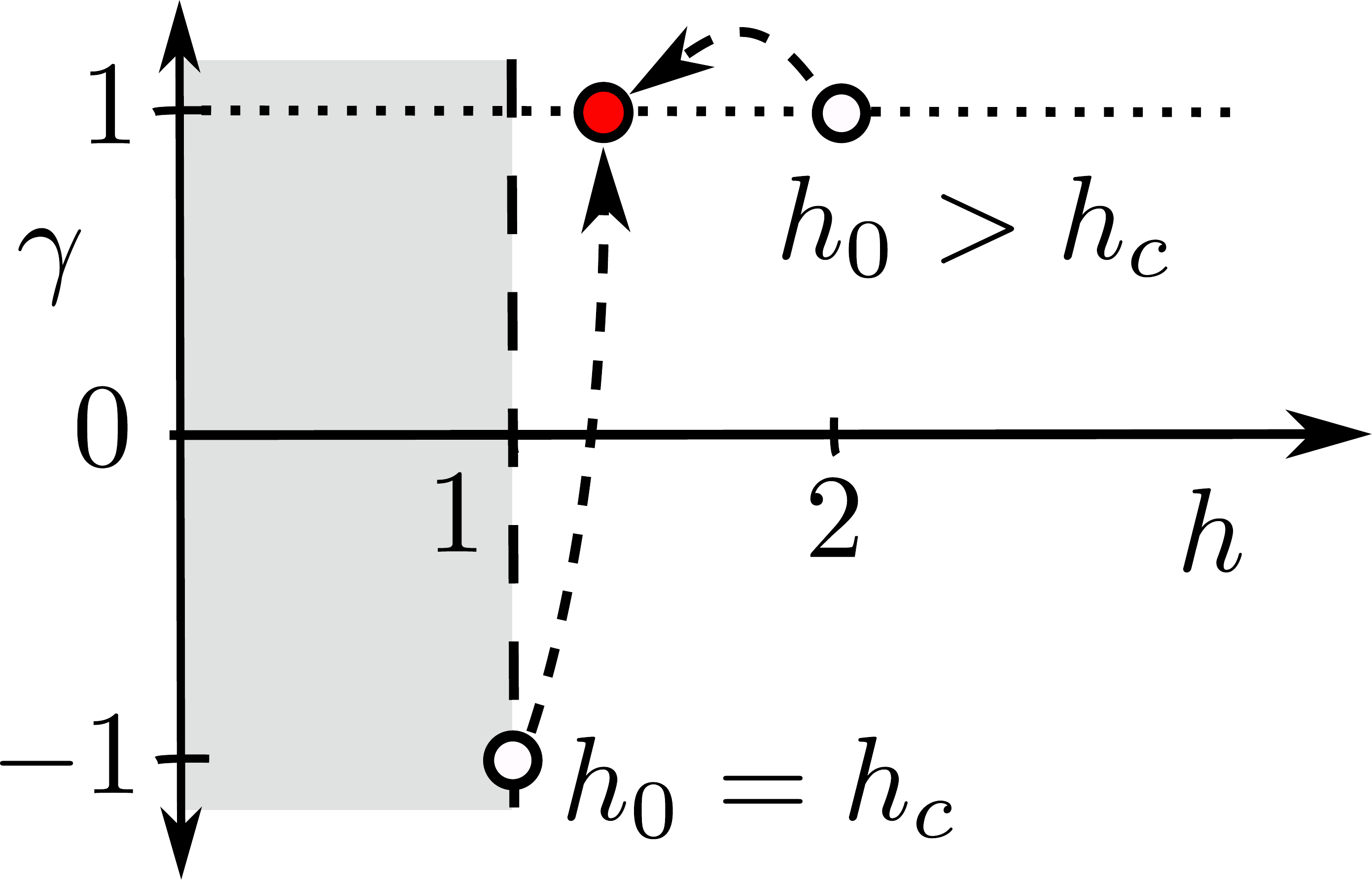}
		\\\vspace{0.2cm}
		\includegraphics*[width=0.991\linewidth]{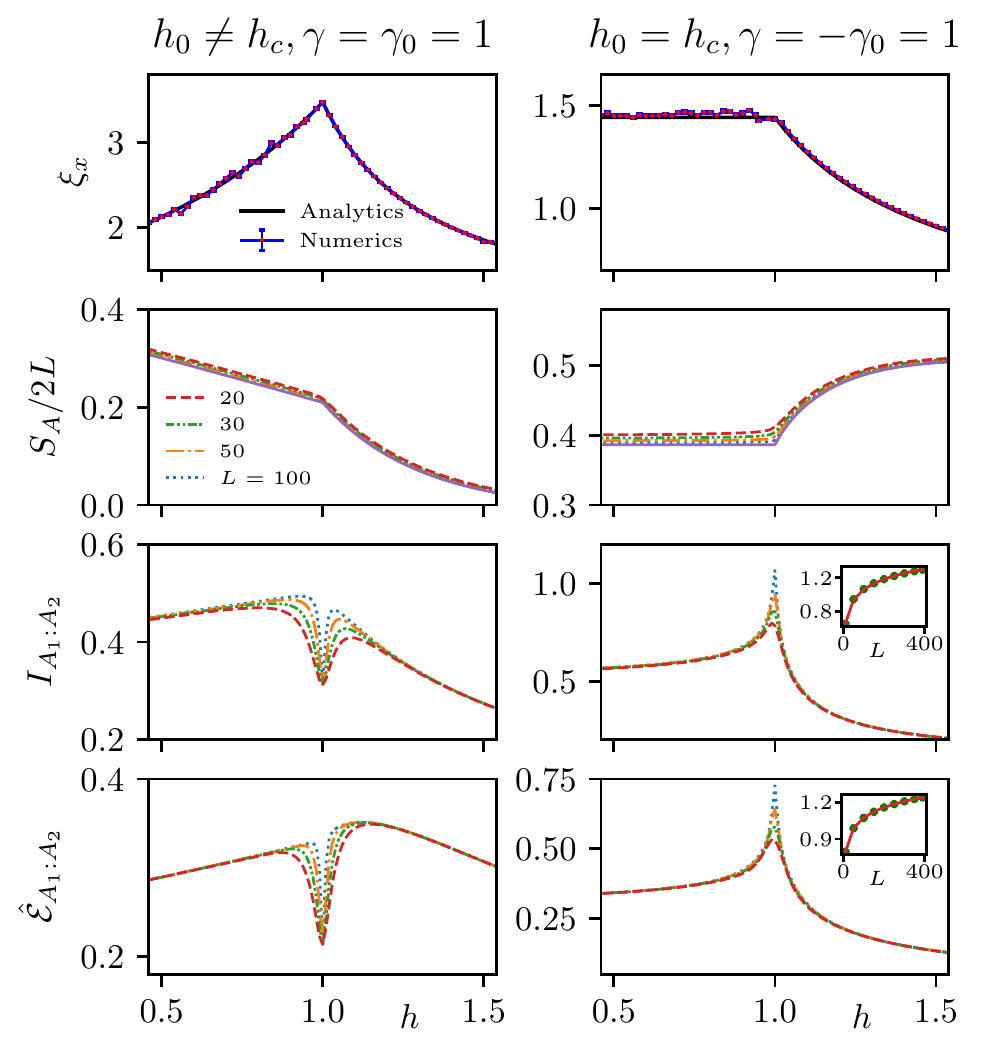}\\\vspace{0.2cm}
		\hspace{1cm}\includegraphics*[width=0.8\linewidth]{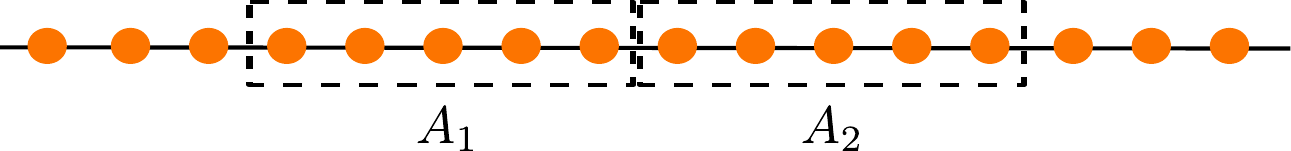}
		\begin{picture}(0,0)
		\put(-195,267){(a)}
		\put(-83,267){(b)}
		\put(-195,208){(c)}
		\put(-83,208){(d)}
		\put(-195,149){(e)}
		\put(-83,149){(f)}
		\put(-195,91){(g)}
		\put(-83,91){(h)}
		\end{picture}	
		\caption{(Top) Ground-state phase diagram of the XY model as a function of $h$ the transverse field and $\gamma$ the anisotropy factor; $h=h_c \equiv 1$ defines the critical line. The arrows schematically represent two quench protocols starting from a noncritical ($h_0>h_c$) or a critical ($h_0=h_c$) initial state. 
			The following quantities are plotted in the long-time stationary state as a function of $h$ along the horizontal dotted line:
			(a,b) longitudinal correlation length (both analytical and numerical results are provided.); (c,d) von Neumann entropy density $S_{A}/2L$ of a region of size $2L$; (e,f) mutual information $I_{A_1:A_2}$, and  
			(g,h) upper bound on log-negativity $\hat{\mathcal{E}}_{A_1:A_2}$ for two adjacent regions of size $L$; see the schematics.
			Different curves in panels (c-h) correspond to different system sizes; see panel (c). Solid (purple) lines in (c,d) are the analytical prediction in the limit $L\to \infty$.
			Both $I_{A_1:A_2}$ and $\hat{\mathcal{E}}_{A_1:A_2}$ exhibit strong dependence on subsystem size near the critical point ($h=h_c$), but exhibit contrasting behaviors (a sharp dip vs a peak) for $h_0 \ne h_c$ and $h=h_c$; the data in the insets (f,h) are consistent with $I_{A_1:A_2}\sim \frac{1}{6} \ln L$ and $\hat{\mathcal{E}}_{A_1:A_2}\sim \frac{1}{8} \ln L$.
		}
		\label{fig_3} 
		\vspace{-.4cm}
	\end{figure}

	\textit{Model}.---We consider the quench dynamics in the anisotropic XY model given by the Hamiltonian
	\begin{equation}
	H(h,\gamma)=-\sum_j \frac{1+\gamma}{2} \sigma^x_j \sigma^x_{j+1} + \frac{1-\gamma}{2} \sigma^y_j \sigma^y_{j+1} + h \sigma^z_j,
	\label{eq_1}
	\end{equation}
	where $\sigma^\alpha~(\alpha=x,y,z)$ are the Pauli operators.
	Here, $h$ is the transverse field, and $\gamma$ defines an anisotropy parameter.
	The ground-state phase diagram of this model is shown in \cref{fig_3} (top): an order-disorder phase transition occurs at $h_c=1$ for any value of $\gamma$.
	A general quench can be parametrized as a sudden change in the parameters of the Hamiltonian defined in \cref{eq_1}: $(h_0,\gamma_0)\to (h, \gamma)$;  the initial state is the ground state of $H(h_0,\gamma_0)$ and evolves under $H(h,\gamma)$. Without loss of generality, we fix $\gamma=1$ in the post-quench Hamiltonian. 
	Quench dynamics of the XY model has been studied extensively~\cite{Calabrese_2006,Mukherjee_2007,Fagotti_2008,Alba_2009,Campos_2011,Najafi_2018,Najafi_2019}.

	Here, we investigate the critical properties of the stationary state at late times. Indeed, recent works have shown that quantum phase transitions leave their fingerprints on quench dynamics~\cite{Bhattacharyya2015,Roy2017,Titum-2019,Haldar-2021}. 
	However, a quantum quench imparts extensive energy and entropy and generically leads to exponentially decaying correlations, hence a lack of genuine critical behavior. 
	Contrary to this picture, we
	show that, depending on the initial state, the long-time stationary state could in fact exhibit critical behavior. 
	To highlight the role of the initial state, we consider two different initial states: (i)~Noncritical: $(h_0,\gamma_0)=(2,1)$ corresponding to a disordered state, and (ii)~Critical: $(h_0,\gamma_0)=(1,-1)$. The two quench protocols are schematically represented in the top panel of \cref{fig_3}. We study the stationary state 
	near the (post-quench) critical point  $h=h_c$ (dotted line in \cref{fig_3}).
	Left (right) columns in \cref{fig_3} correspond to the first (second) protocol, respectively.

	We first consider correlation functions. In a critical ground state ($\gamma>0$), longitudinal correlation functions fall off algebraically as $\rho^{xx}_l\equiv\langle \sigma^x_j\sigma^x_{j+l}\rangle\propto l^{-1/4}$. Away from criticality, they decay exponentially, $\rho^{xx}_l \sim \exp(-l/\xi_x)$, with $\xi_x$ the correlation length; the latter diverges, $\xi_x\sim 1/|h-h_c|$, as $h\to h_c$.
	In contrast, quench dynamics always leads to a disordered stationary state ($\langle\sigma^x\rangle=0$) and a finite correlation length although the latter features a kink at the critical point;
	see \cref{fig_3}(a,b). 
	Transverse correlations behave differently and will be discussed later. 
	A finite correlation length 
	seems to indicate a lack of criticality. However, we show that the critical behavior is hidden in higher-order correlations that are captured via information-theoretic measures.

	Next, we turn to the von Neumann entropy of a connected block of spins $A$ $(=A_1\cup A_2$) of size $2L$; see \cref{fig_3}(c,d)~\cite{supp}. 
	Regardless of the quench protocol, the entropy of the stationary state obeys a volume law. A logarithmic term, if any, would appear to the subleading order in~\cref{Svn_exp}.
	To this end, we consider the mutual information between two (sub)systems $A_1$ and $A_2$ defined as $I_{A_1:A_2}=S_{A_1}+S_{A_2}-S_{A_1\cup A_2}$; this quantity is a measure of the total amount of correlations. 
	In \cref{fig_3}(e,f), we present the mutual information between the two adjacent regions~\cite{supp}. 
	Both quench protocols are sensitive to the critical point, but exhibit very different trends in the stationary state. For the first protocol [\cref{fig_3}(e)], the mutual information is bounded but exhibits a dip at the critical point, which becomes sharper as $L\to\infty$. Interestingly,  $I_{A_1:A_2}\equiv I(L,h)$ 
	displays a finite-size scaling near $h=h_c$:
	\begin{equation}\label{eq:I_scaling}
	I(L, h) =I(\infty, h)-   {\cal F}\big[(h-1)L\big],
	\end{equation}
	with ${\cal F}(x)$ a scaling function~\cite{supp}.
	In contrast, for the second protocol, the mutual information diverges as $I\sim \frac{1}{6}\ln L$ at the critical point (just like the critical ground state); see the inset in \cref{fig_3}(f).  
	Translated to the entanglement entropy, this means that a critical-to-critical quench 
	results in both volume and logarithmic terms in \cref{Svn_exp}. 
	Such logarithmic scaling  
	is indicative of criticality~\cite{Singh2011,Wilms_2012,Alcaraz2014, Stephan2014}, as we further argue below. 
	Away from the critical point and in the limit $L\to \infty$, we find $I\sim -\frac{1}{6}\ln |h-h_c|$~\cite{supp}. Mimicking the ground state, we define a ``mutual information correlation length'' $\xi_{\rm MI}$ as $I\sim \frac{1}{6}\ln \xi_{\rm MI}$ \footnote{This notion of the mutual information correlation length is different from  Ref.~\cite{Wolf_2008}.}.
	We then conclude that $\xi_{\rm MI} \sim 1/|h-h_c|$ diverges in the stationary state  although $\xi_x\sim {\cal O}(1)$.  
	Indeed, mutual information  
	does not overlook any hidden correlations
	which could be invisible to two-point correlations, a property that could be useful for quantum data hiding \cite{Hayden04,Ben-Aroya_2007,Wolf_2008,Hastings_2007}.

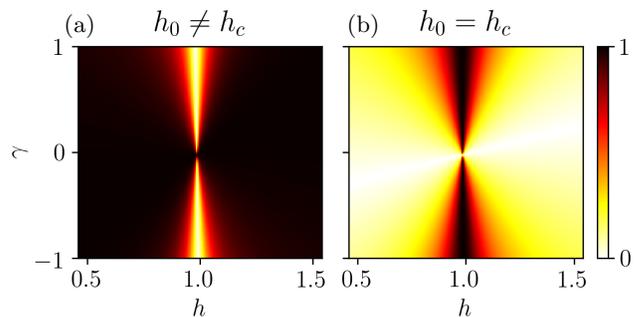
\begin{figure}
		\includegraphics[width=\linewidth]{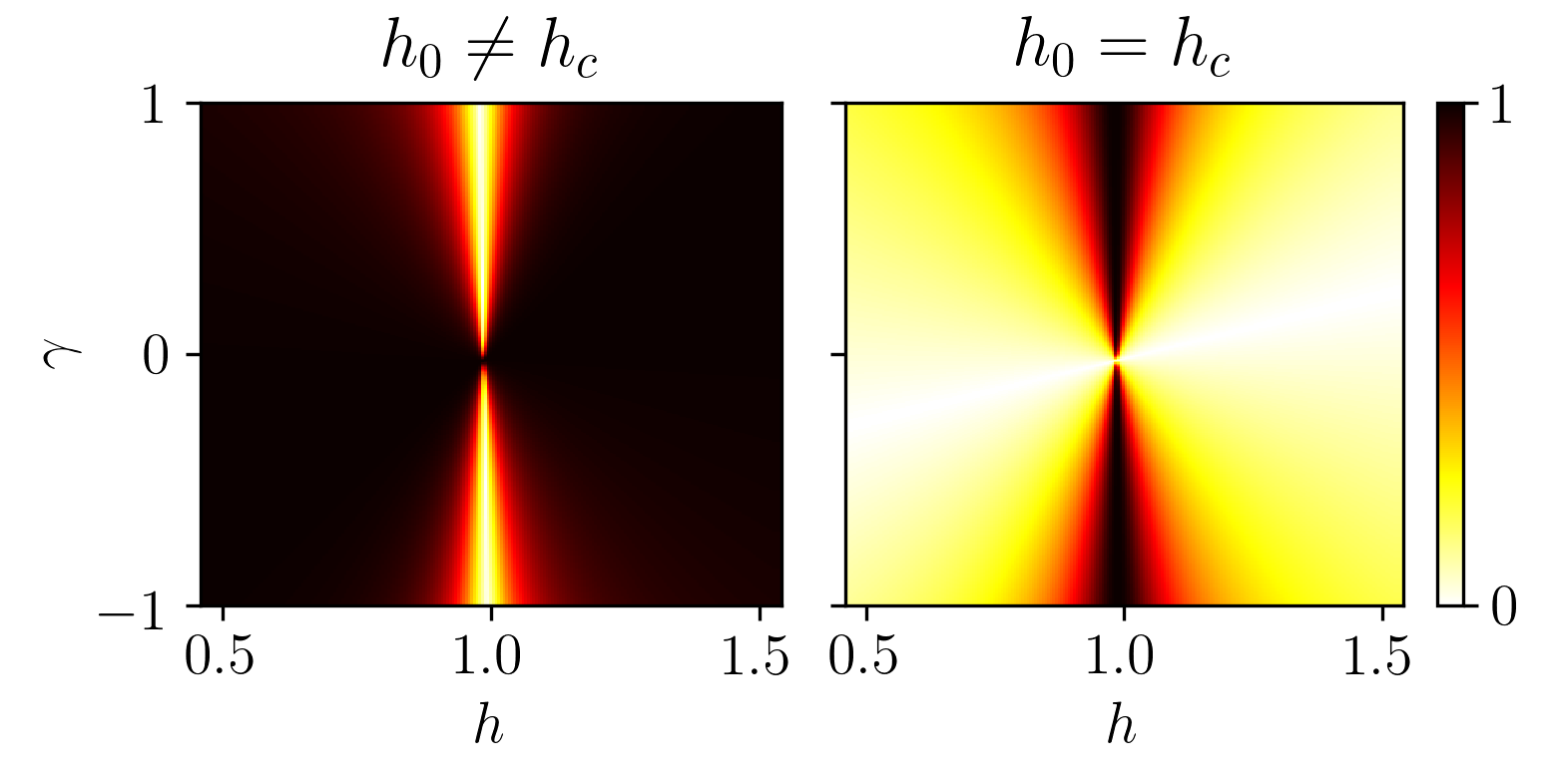}
		\begin{picture}(0,0)
		\put(-95,124.5){(a)}
		\put (12,124.5){(b)}
		\end{picture}
		\vspace{-0.5cm}
		\caption{Effective temperature of the soft mode. The density plot shows $|\tanh(\epsilon_k/2T_k)|=|2n_k-1|$ at $k=0.0125$. The dark (bright) color represents ``low (high) temperatures''. In a quench to the critical point ($h=h_c$), the soft mode (a) overheats when $h_0\ne h_c$, but (b) remains at zero temperature when $h_0=h_c$. In each case, $h_0,\gamma_0$ are the same as Fig.~\ref{fig_3}.}
		\label{fig_2}
		\vspace{-.5cm}
	\end{figure}

	Next, we address the  (classical vs quantum) nature of correlations.
	To this end, we consider an entanglement monotone known as the logarithmic (log-)negativity  
	defined as $\mathcal{E}_{A_1:A_2}=\ln \text{Tr}|\rho_{A}^{T_2}|$ where $A= A_1 \cup A_2$, 
	and $T_2$ represents partial transposition with respect to  
	$A_2$~\cite{Vidal_2002,Plenio_2005}. 
	For technical reasons,
	we calculate a relatively tight upper bound $\hat{\mathcal{E}}$ ($\geq \mathcal{E}$) on log-negativity~\cite{Eisert_2018,Eisler_2015,supp}.  
	In \cref{fig_2}(g,h), we see that $\hat{\mathcal{E}}$ behaves similarly to the mutual information.  
	In particular, $\hat{\mathcal{E}}\sim \frac{1}{8}\ln L$ grows logarithmically with $L$ in the critical-to-critical quench; see the inset in \cref{fig_3}(h). We conclude that the correlations captured by the mutual information are indeed quantum in nature. 
	
	\textit{Fermionic picture.---}%
	To find an insight into our results, we consider an exact mapping to free fermions via the Jordan-Wigner transformation $c_j=\left(\prod_{m<j}\sigma_m^z\right)(\sigma_j^x-i\sigma_j^y)/2$~\cite{supp}.
	This fermionic Hamiltonian can be diagonalized by a Bogoliubov transformation in momentum space, $\eta_k=\cos \frac{\theta_k}{2} c_k + i \sin \frac{\theta_k}{2}c^\dagger_{-k}$, with $e^{-i\theta_k}=(e^{-ik}-h)/\epsilon_k$ and $\epsilon_k=\sqrt{\left(h-\cos k\right)^2+\gamma^2 \sin^2 k}$ the dispersion relation.
	Now, the occupation number of the Bogoliubov modes in the stationary states, $n_k=\langle \eta_k^\dagger \eta_k\rangle$, can be characterized by a $k$-dependent temperature as $\tanh(\epsilon_k /2T_k)\equiv 1-2n_k$, mimicking the Fermi-Dirac distribution for each mode; this $k$ dependence underscores the non-equilibrium nature of the stationary state. 
	Of particular importance is the soft ($k=0$) mode near criticality, $h\to h_c$, where the corresponding energy is vanishing. 
	In \cref{fig_2}, we plot $|2n_k-1|$ with $k\approx 0$ for a range of post-quench parameters $h,\gamma$ and both quench protocols. 
	The plotted quantity vanishes at ``high temperatures'' when $T_k/\epsilon_k\to \infty$ corresponding to a completely mixed state, $n_k=1/2$, and becomes 1 at ``low temperatures'' when $T_k/\epsilon_k\to  0^{\pm}$ corresponding to $n_k=0,1$, respectively.
	For the first protocol [\cref{fig_2}(a)], we see that the soft mode is effectively at infinite temperature when $h\approx h_c$. In contrast, in a critical-to-critical quench, the soft mode is at zero temperature;
	see \cref{fig_2}(b). 
	A vanishing effective temperature in the latter type of quench has been pointed out in the context of mean-field models~\cite{Maghrebi2017arxiv,Chiocchetta2017,Titum_2020}.
	We remark that other modes at  
	$k\ne 0$ are generically at a nonzero temperature $T_k \ne 0$, regardless of the type of quench resulting in the extensive entropy of the stationary state~\cite{supp}.
	
	The distinctive behavior of the soft mode leaves its fingerprint on long-distance fermionic correlations in the stationary state.
	The latter are fully described by a single correlator in terms of Majorana fermions, $g^{\rm st}_l\equiv i\langle a^x_j a^y_{j+l}\rangle$, where $a^{x}=c_j+c_j^\dagger$ and $a^y=i(c_j^\dagger - c_j)$~\cite{supp}.
	For a critical-to-critical quench, 
	this correlator takes the form
	\begin{equation}\label{eq:g-l}
	g_l^{\rm st}
	=\frac{2 + 4 l}{-3\pi + 4\pi l ( l+1)} \xrightarrow[l\to \infty]{} \frac{1}{\pi l}\,. 
	\end{equation}
	This should be contrasted with the 
	critical ground state 
	where $g_l^{\rm gs} =-2/(\pi+2\pi l)\rightarrow -1/(\pi l)$ as $l\to \infty$. Despite the deviation at short distances, the asymptotic behavior of the correlation function 
	is identical to that of the ground state up to a sign. Note that in the stationary state, $n_{k=0}=1$ corresponding to vanishing, but negative effective temperature [see~\cref{fig_2}(b)], which is responsible for quantum criticality and the overall sign in $g_l^{\rm st}$. 
	In contrast, in a noncritical-to-critical quench [\cref{fig_2}(a)], $g_l^{\rm st}$ decays exponentially, consistent with a high-temperature state.

	The above analysis suggests that fermionic observables that are not sensitive to short-wavelengths must exhibit the same scaling as the critical ground state. Besides correlation functions, mutual information too
	is independent of the ultraviolet cutoff~\cite{Eisert_2010} and instead captures long-range correlations
	\footnote{In contrast with the mutual information, the entanglement entropy generally diverges with the ultraviolet cutoff.}. 
	Let us first consider the mutual information $I^f_{A_1:A_2}$ corresponding to the fermionic lattice model. Given that the model is Gaussian, and that it exhibits the same long-range correlations as the critical ground state [\cref{eq:g-l}], we conclude that $I^f_{A_1:A_2}\sim \frac{1}{6}\ln L$.
	We remark that highly-excited states of free fermions can be constructed where mutual information scales logarithmically~\cite{Alba_2009, Ares_2014}. Here, we have shown that such behavior emerges naturally in a critical-to-critical quench. 
	For the spin model, note that $I^f_{A_1:A_2}= I_{A_1:A_2}$ for adjacent regions since the corresponding spin operators can be written in terms of fermions in the same region~\cite{Vidal_2004}. It follows that $I_{A_1:A_2}\sim \frac{1}{6}\ln L$ for the critical-to-critical quench, consistent with our numerics.
	A remark is in order. Our results are consistent with CFT calculations where it was shown that the entanglement of the initial state survives in the dynamics~\cite{Cardy-2016}; however, this conclusion does not hold in a quench to a noncritical point even if the initial state is critical and highly entangled [see \cref{fig_3}(f,h)]. 
	
	Next, we explain the features observed in the mutual information and log-negativity for the noncritical-to-critical quench [\cref{fig_3}(e,g)]. As shown in
	\cref{fig_2}(a), this quench exhibits a stark discontinuity in the effective temperature: $T_k/\epsilon_k$ diverges exactly at $h=h_c$ but vanishes for any $h\ne h_c$. The \textit{overheating} of the soft mode at the critical point ($h=h_c$) is responsible for the sharp dip observed in \cref{fig_3}(e,g). Furthermore, the finite-size scaling with $(h-1)L$ in \cref{eq:I_scaling} reflects the scaling behavior of the occupation number, $n_k\sim {\cal N}\big(k/(h-1)\big)$~\cite{supp}.

	\begin{figure}[htbt]
		\hspace{1.2cm}\includegraphics[width=0.8\linewidth]{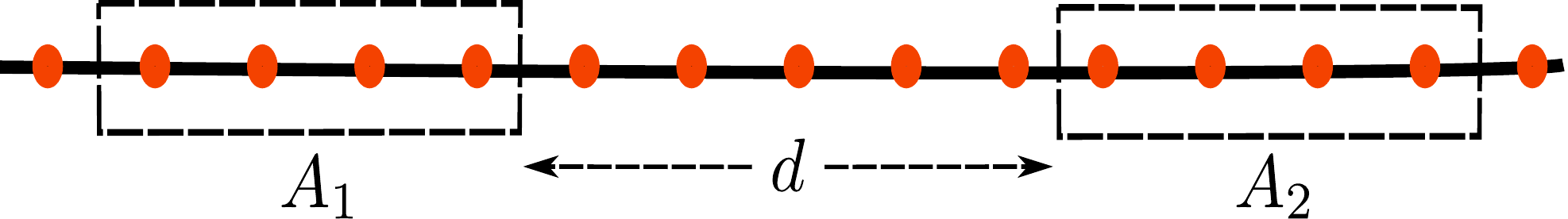}\\\vspace{0.1cm}
		\includegraphics[width=\linewidth]{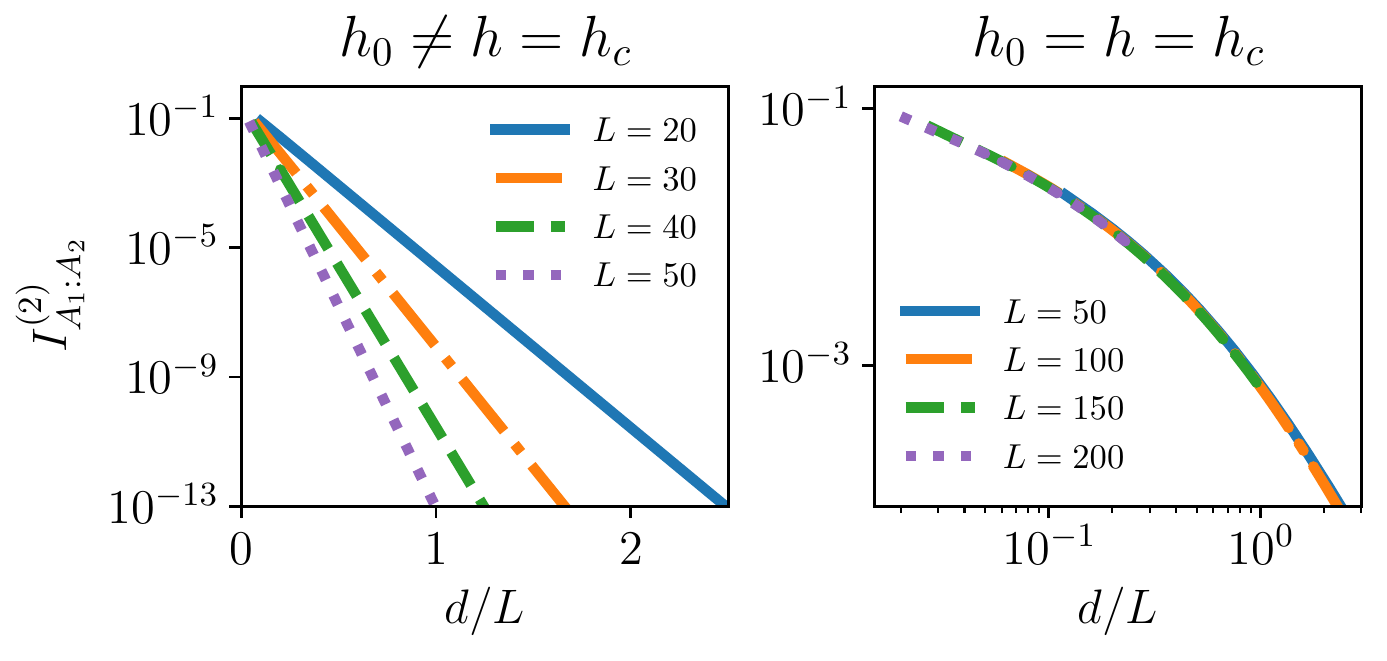}
		\begin{picture}(0,0)
		\put(-80,118){(a)}
		\put (32,118){(b)}
		\end{picture}
		\vspace{-0.4cm}
		\caption{Scaling behavior of disjoint R\'{e}nyi information $I^{(2)}_{A_1:A_2}$ in the stationary state of quench dynamics. Schematics represent two disjoint regions $A_1$ and $A_2$ each of length $L$ separated by a distance $d$. Disjoint R\'{e}nyi information $I_2$ is plotted for (a) noncritical-to-critical quench,
			and (b) critical-to-critical quench.
			The latter exhibits scaling invariance, indicative of criticality. In each case, $h_0,\gamma_0,\gamma$ are the same as \cref{fig_3}.}
		\label{fig_4}
		\vspace{-.4cm}
	\end{figure}
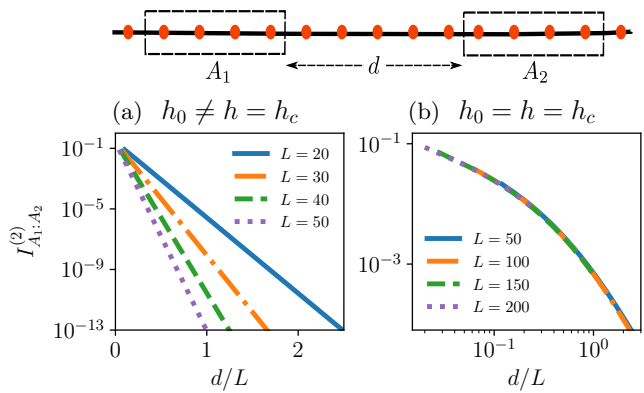

	\textit{Hidden criticality.---}%
	The critical nature of the model, while manifest in terms of fermions, becomes hidden when examined through spin correlations. 
	This surprising feature should be attributed to the Jordan-Wigner string operator. 
	On the other hand, the connected transverse spin correlations are directly determined from fermionic correlations; in the critical-to-critical quench, they scale as $\langle \sigma^z_{j+l} \sigma^z_{j}\rangle_c=-g^{\rm st}_{l+1}g^{\rm st}_{1-l}\sim
	1/ \pi^2 l^2$  
	as $l\rightarrow \infty$, identical to the critical ground state. 
	In spite of this, we argue that the universal logarithm in the information measures cannot be attributed just to the algebraic decay of transverse correlations.
	First, if long-range correlations only involve $\sigma^z$ operators, they will be of a classical nature, 
	but this would be incompatible with the logarithmic divergence of log-negativity. 
	Furthermore, two-point correlations alone cannot violate the area law for the mutual information if the decay is faster than $1/l$~\cite{supp}.
	In contrast with transverse correlations, the fermionic two-point correlations in \cref{eq:g-l} decay as $1/l$, which, in the spin language, translates to a string operator:
	\begin{equation}\label{eq:multi-point_correlations}
	\Big\langle \sigma^x_j \sigma^x_{j+l} \prod_{j< m<j+l} 
	\sigma^z_m \Big\rangle \sim -\frac{1}{\pi l}.
	\end{equation} 
	This shows explicitly that criticality is hidden in higher-order spin correlations.

	\textit{Scaling invariance of disjoint R\'{e}nyi mutual information}.---%
	A direct signature of criticality is scale invariance due to the divergence of the correlation length. While correlation functions decay exponentially, information-theoretic measures are a suitable candidate to display such scale invariance.
	We inspect the mutual information between two disjoint regions of size $L$ separated by a distance $d$. For technical reasons, we consider the R\'{e}nyi mutual information $I^{(\alpha)}_{A_1:A_2}$ defined analogously from the R\'{e}nyi entropy, $R^{(\alpha)}_{A}=\frac{1}{1-\alpha}\ln \text{Tr}_A \rho_A^{\alpha}$ and take $\alpha=2$ for simplicity.
	Critical behavior, if any, dictates 
	\begin{equation}
	I^{(2)}_{A_1:A_2} = {\mathcal I}(d/L),    
	\end{equation}
	with $\mathcal I$ a scaling function which only depends on the ratio $d/L$, independent of any intrinsic scales.
	In \cref{fig_4}, 
	we show that the noncritical-to-critical quench shows no such scaling, but the critical-to-critical quench is manifestly scale invariant. 
	This provides further evidence that the latter quench leads to genuinely critical behavior \cite{supp}.

	\textit{Experimental realization.---}%
	The quench experiments studied here may be experimentally realized in a variety of quantum simulator platforms. Particularly, we envision 1D arrays of Rydberg atoms trapped using optical tweezers as the ideal platform to study these Hamiltonians~\cite{Browaeys_NatPhys2020}. These systems have long coherence times, tunable interactions, and have been used to implement a variety of spin models as well as universal quantum computing~\cite{Morgado_2021}. 
	To investigate the hidden criticality in the experiment, a challenging aspect is the preparation of a critical state which may be possible using variational algorithms~\cite{Pagano_PNAS2020}. The measurement of entanglement, such as the R\'{e}nyi entropy of a quantum state, can be performed using the statistical correlations in randomized measurements~\cite{Elben_PRA2019,Roos_Science2019}.
	Finally, we note that while the results in this paper have focused on the stationary state of the quench dynamics, we expect the essential features of the hidden criticality, especially its scaling behavior, to be manifest in the intermediate time dynamics.

	\textit{Conclusion and outlook.---}%
	We have studied the critical behavior in the long-time stationary state of an integrable spin chain upon a sudden quench. We have shown that, for certain quenches, the stationary state exhibits quantum critical behavior which cannot be detected through the local order parameter and is instead hidden in higher-order correlations, which we identify through information-theoretic measures. Our findings open up a new frontier for investigating quantum criticality in quench dynamics beyond the ground-state order-disorder phase transitions. An immediate direction is to investigate the distinctive features at intermediate dynamics. 
	Furthermore, it is worthwhile to study interacting integrable spin models. 
	A particularly interesting direction is to investigate if chaotic Hamiltonians leave similar fingerprints at intermediate times \cite{Haldar-2021}. Finally, exploring connections with data hiding in quantum information is worthwhile \cite{Hayden04,Ben-Aroya_2007}.

	
	\begin{acknowledgments}
		We thank M.A.~Rajabpour, A.~Polkovnikov, M.~Rigol, and H.~Shapourian
		for insightful discussions.
		M.F.M. acknowledges support from NSF under Grant No. DMR-1912799. S.P. is supported by the start-up funding from Michigan State University.
		P.T. acknowledges funding from the U.S. Department of Energy (DOE), Office of Science, Office of Advanced Scientific Computing Research (ASCR) Quantum Computing Application Teams program, under fieldwork proposal number ERKJ347. This research was supported in part by the National Science Foundation under Grant No. NSF PHY-1748958.
	\end{acknowledgments}
	
	
	\bibliographystyle{apsrev4-1}
	\bibliography{paper} 

	

\section*{Supplementary material}

# Supplemental Material for
# "Hidden Quantum Criticality and Entanglement in Quench Dynamics"

Sanku Paul,[1,2] Paraj Titum,[3,2,4] and Mohammad F. Maghrebi[1,2]

[1]*Department of Physics and Astronomy, Michigan State University, East Lansing, Michigan 48824, USA*
[2]*Kavli Institute for Theoretical Physics, University of California, Santa Barbara, CA 93106-4030, USA*
[3]*Johns Hopkins University Applied Physics Laboratory, Laurel, Maryland 20723, USA*
[4]*Joint Quantum Institute, NIST/University of Maryland, College Park, Maryland 20742, USA*

In this Supplemental Material, we provide additional details for the results stated in the main text. In Sec. S.I, we provide the mapping between spins and fermions through the Jordan-Wigner transformation, report the two-point correlation functions both for spins and fermions, and further discuss the properties of the fermionic occupation number for different types of quench. In Sec. S.II, we provide details on how different measures of entanglement as well as information, including von-Neumann entropy, mutual information, an upper bound on logarithmic negativity, and disjoint Rényi mutual information can be described via the fermionic model. In Sec. S.III, we discuss the finite-size scaling of the mutual information and log-negativity near the critical point. Finally, in Sec. S.IV, we argue that two-point correlations alone, if decaying faster than $1/l^\alpha$ with $\alpha > 1$, cannot lead to a (logarithmic) violation of the area law.

## S.I. XY MODEL : FERMIONIC PICTURE

### A. Jordan-Wigner transformation

In this section, we discuss the mapping of the transverse-field XY model introduced in Eq. (2) of the main text to a non-interacting fermionic model. This mapping is facilitated via a Jordan-Wigner transformation given by [S1]

$$c_j = \left(\prod_{m<j} \sigma_m^z\right) \frac{\sigma_j^x - i\sigma_j^y}{2},$$
$$c_j^\dagger = \left(\prod_{m<j} \sigma_m^z\right) \frac{\sigma_j^x + i\sigma_j^y}{2}, \quad \text{(S.1)}$$

where the fermionic operators satisfy the anticommutation relations $\{c_j, c_m\} = 0$ and $\{c_j^\dagger, c_m\} = \delta_{jm}$. This provides a non-local transformation as it maps a string of the first $j$ spins to a fermion at site $j$. The resulting fermionic Hamiltonian becomes a quadratic fermionic model with nearest-neighbor interactions and is expressed as

$$H = -\sum_j \left[c_j^\dagger c_{j+1} + \gamma c_j^\dagger c_{j+1}^\dagger + \text{h.c.}\right] + 2h c_j^\dagger c_j. \quad \text{(S.2)}$$

The Hamiltonian in Eq. (S.2) can be diagonalized by first going to the Fourier space and introducing the Bogoliubov transformation given by [S1]

$$\eta_k = \cos(\theta_k/2)\, c_k + i\sin(\theta_k/2)\, c_{-k}^\dagger,$$
$$\eta_k^\dagger = \cos(\theta_k/2)\, c_k^\dagger - i\sin(\theta_k/2)\, c_{-k}, \quad \text{(S.3)}$$

where the phase $\theta_k$ is given by

$$e^{-i\theta_k} = \frac{e^{-ik} - h}{\epsilon_k}. \quad \text{(S.4)}$$

The resulting Hamiltonian becomes non-interacting in the fermionic basis and takes the form [S1]:

$$H = \sum_k \epsilon_k (\eta_k^\dagger \eta_k - \frac{1}{2}), \quad \text{(S.5)}$$

with the dispersion

$$\epsilon_k = \sqrt{(h - \cos k)^2 + \gamma^2 \sin^2 k}\,. \tag{S.6}$$

It is further useful to represent the Hamiltonian in Eq. (2) in terms of Majorana fermionic operators which are obtained as [S2]

$$a_j^x = c_j + c_j^\dagger, \qquad a_j^y = i(c_j^\dagger - c_j)\,. \tag{S.7}$$

These Majorana operators satisfy the anti-commutation relation $\{a_j^\alpha, a_m^\beta\} = 2\delta_{jm}^{\alpha\beta}$, where $\alpha, \beta = x, y$. In this basis, the Hamiltonian in Eq. (2) takes the form

$$H = i \sum_j \left( \frac{1+\gamma}{2} a_j^x a_{j+1}^y - \frac{1-\gamma}{2} a_j^y a_{j+1}^x + h a_j^y a_j^x \right). \tag{S.8}$$

### B. Two-point correlation functions

The fermionic Hamiltonian in Eq. (S.8) being quadratic, the density matrix of the system becomes Gaussian [S2], and can be completely characterized in terms of the two-point correlation functions of Majorana fermions:

$$\langle a_j^\alpha a_m^\beta \rangle = \delta^{\alpha\beta}\delta_{jm} + i\Gamma_l^{\alpha\beta}, \qquad l = m - j\,, \tag{S.9}$$

with the correlation matrix $\Gamma_l^{\alpha\beta}$. For a connected spatial region $A$ of size $L_A$, we introduce the correlation matrix $\Gamma^A$ as [S1]

$$\Gamma^A = \begin{bmatrix} \Gamma_0 & \Gamma_{-1} & \cdots & \Gamma_{1-L_A} \\ \Gamma_1 & \Gamma_0 & \cdots & \cdot \\ \vdots & \vdots & \ddots & \vdots \\ \Gamma_{L_A-1} & \cdot & \cdots & \Gamma_0 \end{bmatrix}, \tag{S.10}$$

with

$$\Gamma_l = \begin{bmatrix} -f_l & g_l \\ -g_{-l} & f_l \end{bmatrix}, \tag{S.11}$$

where the two-point correlation functions are given by $f_l + i\delta_{l0} = i\langle a_j^x a_{j+l}^x \rangle = i\langle a_{j+l}^y a_j^y \rangle$ and $g_l = i\langle a_j^x a_{j+l}^y \rangle$. The two-point spin correlations can be obtained from those of Majorana fermions. Specifically, we have [S1]

$$\rho^{xx}(l) = \langle \sigma_j^x \sigma_{j+l}^x \rangle = \text{Pf}(\Gamma^A)\,, \tag{S.12}$$

$$\rho_c^{zz}(l) = \langle \sigma_{j+l}^z \sigma_j^z \rangle - \langle \sigma_j^z \rangle^2 = f_{-l}f_l - g_{l+1}g_{1-l}\,, \tag{S.13}$$

where $\rho^{xx(yy)}(l) \equiv \langle \sigma_j^{x(y)} \sigma_{j+l}^{x(y)} \rangle$ are the longitudinal correlation functions, $\rho_c^{zz}(l)$ is the connected transverse correlation function, and the symbol Pf denotes the Pfaffian.

We shall consider the thermodynamic limit, i.e., an infinitely long chain, and a general quench of the form $\{h_0, \gamma_0\} \to \{h, \gamma\}$. The time-dependent correlation functions $g_l(t)$ and $f_l(t)$ can be obtained from a modified Bogoliubov angle defined by $\Delta_k = \theta_k - \theta_k^0$, where $\theta_k$ ($\theta_k^0$) defined in Eq. (S.4) corresponds to the post- (pre-)quench Hamiltonian. Explicitly, we have [S3]

$$f_l(t) = \frac{i}{2\pi} \int_0^{2\pi} dk \ e^{-ilk} \sin \Delta_k \sin 2\epsilon_k t\,, \tag{S.14}$$

$$g_l(t) = \frac{1}{2\pi} \int_0^{2\pi} dk \ e^{-ilk} e^{-i\theta_k} (\cos \Delta_k - i \sin k \cos 2\epsilon_k t)\,, \tag{S.15}$$

where

$$\cos \Delta_k = \frac{hh_0 - (h + h_0)\cos k + \cos^2 k + \gamma\gamma_0 \sin^2 k}{\epsilon_k \epsilon_k^0}\,,$$

$$\sin \Delta_k = -\sin k \frac{\gamma h_0 - \gamma_0 h - (\gamma - \gamma_0)\cos k}{\epsilon_k \epsilon_k^0}\,, \tag{S.16}$$

and $\epsilon_k$ ($\epsilon_k^0$) is the dispersion of the fermionic excitation given in Eq. (S.6) corresponding to the post- (pre-)quench Hamiltonian. By replacing $h_0 = h$ and $\gamma_0 = \gamma$, one can obtain the correlation functions for the ground state.

As we are interested in the properties of the stationary state, we shall focus on the asymptotic behavior in the limit $t \to \infty$. In this limit, the time-dependent sine and cosine functions in Eqs. (S.14) and (S.15) become highly oscillatory and can be neglected. The correlation functions $f_l^{\rm st} \equiv f_l(t \to \infty)$ and $g_l^{\rm st} \equiv g_l(t \to \infty)$ of the Majorana operators then become

$$f_l^{\rm st} = 0, \qquad g_l^{\rm st} = \frac{1}{2\pi} \int_0^{2\pi} dk \ e^{-ilk} e^{-i\theta_k} \cos \Delta_k \equiv \frac{1}{2\pi} \int_0^{2\pi} dk \ e^{-ikl} g_k^{\rm st}. \tag{S.17}$$

Here, we have defined the correlation function in the momentum space $g_k^{\rm st} = e^{-i\theta_k} \cos \Delta_k$.

The occupation number too can be computed as

$$n_k = \langle \eta_k^\dagger \eta_k \rangle = \frac{1}{2}(1 - \cos \Delta_k). \tag{S.18}$$

Notice that for the ground state ($h_0 = h$ and $\gamma_0 = \gamma$), we have $\cos \Delta_k = 1$ and thus $n_k = 0$.

In the subsequent subsections, we report the analytical results for the two-point correlation functions and the fermionic occupation number for the ground state and different quench protocols considered in the main text.

1. *Critical ground state ($h_0 = h = h_c$ and $\gamma = \gamma_0 \neq 0$)*

- The correlation function $g_l^{\rm gs}$ at long distances $l \gg 1$ can be inferred from the behavior of the correlation function at small wavevectors, $k \to 0$. Notice that in the ground state (critical or not), we have $\cos \Delta_k = 1$ consistent with the occupation number $n_k = 0$. We find

$$g_k^{\rm gs} \sim -\frac{i}{|\gamma|} \text{sgn} k, \tag{S.19a}$$

$$g_l^{\rm gs} \sim -\frac{1}{\pi|\gamma|} \frac{1}{l}. \tag{S.19b}$$

Notice that the non-analyticity in the limit $k \to 0$ results in an algebraic decay of correlations at long distances. For the special case where $\gamma = 1$, the function $g_l^{\rm gs}$ can be computed exactly as [S4]

$$g_l^{\rm gs} = -\frac{1}{\pi(l + 1/2)}, \tag{S.20}$$

which reduces to Eq. (S.19b) when $|l| \gg 1$.

Finally, for the highest excited state, we have $\cos \Delta = -1$ corresponding to fully occupied states ($n_k = 1$), where we find $g^{\rm ex} = -g^{\rm gs}$, therefore the highest excited state is only different from that of the ground state by an overall sign.

- Two-point spin correlation functions at long distances $|l| \gg 1$ and for $\gamma > 0$ are given by [S1]

$$\rho^{xx}(l) \propto l^{-1/4}, \quad \rho^{yy}(l) \propto l^{-9/4}, \quad \rho_c^{zz}(l) \sim \frac{1}{\pi^2 \gamma^2 l^2}. \tag{S.21}$$

For $\gamma < 0$, the scaling of $\rho^{xx}$ and $\rho^{yy}$ is swapped. All correlation functions decay algebraically, which is indicative of criticality.

- For the ground state, the fermionic occupation number is simply $n_k = 0$.

2. *Critical-to-critical quench ($h_0 = h = h_c, \gamma_0 \neq \gamma$)*

- At small wavevectors, $k \to 0$, and long distances $|l| \gg 1$, we find

$$g_k^{\rm st} \sim -\frac{i \text{sgn}(\gamma_0)}{\gamma} \text{sgn}(k), \tag{S.22a}$$

$$g_l^{\rm st} \sim \frac{\text{sgn}(\gamma_0)}{\pi \gamma} \frac{1}{l}. \tag{S.22b}$$

Similar to a critical ground state, the two-point correlation functions decay inversely with the distance in spite of a quench. For the special case where $-\gamma_0 = \gamma = 1$, the function $g_l^{\text{st}}$ can be computed exactly as (reported from Eq. (4) of the main text for completeness)

$$g_l^{\text{st}} = \frac{2 + 4l}{-3\pi + 4\pi l(l+1)}, \tag{S.23}$$

which reduces to Eq. (S.22b) when $|l| \gg 1$.

- Two-point spin correlation functions at long distances $|l| \gg 1$ are given by

$$\rho^{xx}(l) \propto e^{-l/\xi_x}, \quad \rho^{yy}(l) \propto e^{-l/\xi_y}, \quad \rho_c^{zz}(l) \sim \frac{1}{\pi^2 \gamma^2 l^2}, \tag{S.24}$$

where $\xi_{x,y}$ are the respective correlation lengths. For $-\gamma_0 = \gamma = 1$, the exponential decay of longitudinal correlations are confirmed numerically in Fig. 1 of the main text; the corresponding correlation lengths are computed analytically in Ref. [S5]. However, the connected transverse correlation decay algebraically similar to a critical ground state.

- As we are mainly interested in the excitation of the soft mode $k \to 0$, we report the fermionic occupation number in this limit:

$$n_{k=0} = \frac{1}{2}[1 - \text{sgn}(\gamma)\text{sgn}(\gamma_0)]. \tag{S.25}$$

Depending on the sign of $\gamma$ and $\gamma_0$, the soft mode $n_{k=0}$ is either unexcited (0) or fully occupied 1, corresponding to the ground state and the highly excited state, respectively. Both states exhibit criticality at $h = h_c$. For the special case where $-\gamma_0 = \gamma = \pm 1$, we find

$$n_k = \cos^2 \frac{k}{2}. \tag{S.26}$$

In the limit $k \to 0$, we recover the general result in Eq. (S.25).

### 3. Critical-to-noncritical quench ($h_0 = h_c, h > h_0$ and $\gamma_0, \gamma \neq 0$)

- At small wavevectors, $k \to 0$, and long distances $|l| \gg 1$, we find

$$g_k^{\text{st}} \sim \frac{h - 1 + 2\gamma_0\gamma}{2|\gamma_0|(h-1)}|k|, \tag{S.27a}$$

$$g_l^{\text{st}} \sim -\frac{h - 1 + 2\gamma_0\gamma}{2\pi|\gamma_0|(h-1)} \frac{1}{l^2}, \tag{S.27b}$$

The correlation function decays algebraically with the distance as $1/l^2$ but faster compared to either the ground state or the critical-to-critical quench where it decays as $1/l$.

- Two-point spin correlation functions at long distances $|l| \gg 1$ are given by

$$\rho^{xx}(l) \propto e^{-l/\xi_x}, \quad \rho^{yy}(l) \propto e^{-l/\xi_y}, \quad \rho_c^{zz}(l) \sim -\frac{(h - 1 + 2\gamma_0\gamma)^2}{4\pi^2\gamma_0^2(h-1)^2} \frac{1}{l^4}. \tag{S.28}$$

The longitudinal correlation functions decay exponentially (see Fig. 1 and Ref. [S5]), while the connected transverse correlations decay algebraically as $1/l^4$ compared to the $1/l^2$ in the ground state or the critical-to-critical quench.

- At small wavevectors ($|k| \ll 1$), the occupation number becomes

$$n_k \sim \frac{1}{2} - \frac{(h - 1 + 2\gamma_0\gamma)}{4|\gamma_0||h-1|}|k|. \tag{S.29}$$

Therefore, a quench starting from a critical state leads to $n_{k=0} = \frac{1}{2}$ which can be interpreted as the $k=0$ mode heating up to infinite temperature. Furthermore, in the limit $|k|, |h-1| \ll 1$, but for arbitrary $k/(h-1)$, one can write the occupation number $n_k$ in terms of a scaling function $\tilde{\mathcal{N}}$ as

$$n_k = \tilde{\mathcal{N}}\left(\frac{k}{h-1}\right), \quad \text{with} \quad \tilde{\mathcal{N}}(x) = \frac{1}{2} - \frac{1}{2}\frac{\text{sgn}(\gamma_0)\gamma|x|}{\sqrt{1+\gamma^2 x^2}}. \tag{S.30}$$

The scaling form of this equation suggests that, near criticality, relevant physical quantities (including the mutual information and log-negativity) should exhibit a crossover as a function of $(h-1)L$. This is consistent with the crossover from the scaling behavior $\sim b \ln L$ when $(h-1)L \ll 1$ to another given by $\sim -b \ln(h-1)$ for $(h-1)L \gg 1$; the coefficient $b = \frac{1}{6}$ for the mutual information and $b = \frac{1}{8}$ for the log-negativity; see Sec. S.III.

### 4. Noncritical-to-general quench ($h_0 \neq h_c$, and $\gamma_0, \gamma \neq 0$)

- In this case, the function $g_k^{\text{st}}$ is fully analytic in $k$ unlike the other cases discussed above where a non-analyticity appears at $k = 0$. Thus, in this case, all correlations functions decay exponentially in position space:

$$\rho^{xx}(l) \propto e^{-l/\xi_x}, \quad \rho^{yy}(l) \propto e^{-l/\xi_y}, \quad \rho_c^{zz}(l) \propto e^{-l/\xi_z}, \tag{S.31}$$

with $\xi_{x,y,z}$ the respective correlation lengths. Beside the longitudinal correlation lengths reported in Fig. 1 of the main text, the transverse correlation length $\xi_z$ is reported in Fig. S.1; see also Ref. [S5] for analytical expressions.

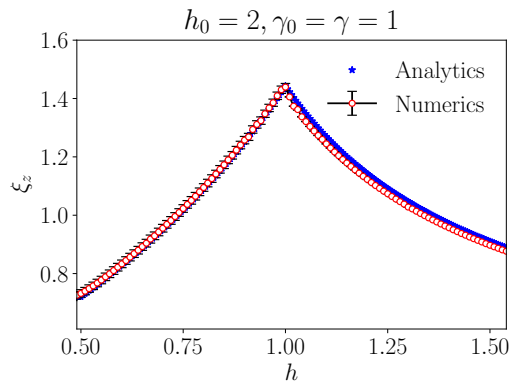

FIG. S.1. Transverse correlation length $\xi_z$ in a quench from a noncritical initial state with $h_0 = 2, \gamma_0 = 1$. In the numerical data, vertical bars signify the one standard deviation error in calculating the correlation length using an exponential fit to the connected transverse correlation in Eq. (S.31). The analytical result is calculated using Eq. (30) in Ref. [S5].

- For a quench starting from a noncritical initial state, the occupation number behaves qualitatively differently depending on the final state. For simplicity, we assume that $h_0 > 1$ and $\gamma_0 \gamma > 0$; the extension to the general case should be straightforward. At small wavevectors ($|k| \ll 1$), we find

$$n_k \sim \begin{cases} 1 - Ak^2, & h < 1, \\ \frac{1}{2} - B|k|, & h = 1, \\ Ak^2, & h > 1, \end{cases} \tag{S.32}$$

where

$$A = \frac{[\gamma_0(1-h) + \gamma(h_0-1)]^2}{4(h-1)^2(h_0-1)^2}, \quad B = \frac{(h-1+2\gamma_0\gamma)}{4|\gamma_0||h-1|}. \tag{S.33}$$

Notice that there is a discontinuity at $h = 1$; see also the right plot of Fig. S.2. Specifically, we observe that for $h \neq 1$, the occupation number $n_{k=0}$ takes an extreme value (0 or 1) while exactly at $h = 1$ we find $n_{k=0} = 1/2$.

Furthermore, in the limit $|k|, |h-1| \ll 1$, but for arbitrary $k/(h-1)$, one can write the occupation number $n_k$ in terms of a scaling function $\mathcal{N}$ as

$$n_k = \mathcal{N}\left(\frac{k}{h-1}\right), \qquad \text{with} \quad \mathcal{N}(x) = \frac{1}{2} - \frac{1}{2}\frac{\text{sgn}(h-1)}{\sqrt{1+\gamma^2 x^2}}. \tag{S.34}$$

The discontinuous behavior of the occupation number $n_k$ as a function of $h$ near the critical point $h = 1$ is responsible for the sharp dip observed in the mutual information and the upper bound on log-negativity in Fig. 1(e,g) of the main text. The scaling form of the above equation accounts for the finite-size scaling of the mutual information and log-negativity; see Sec. S.III.

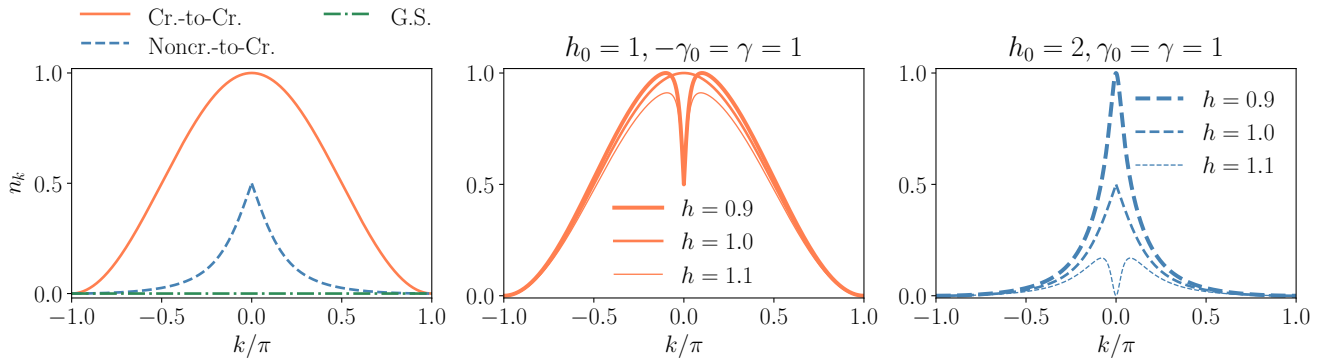

FIG. S.2. Fermionic occupation number $n_k$ for different quench protocols. (Left) Occupation number is shown for the ground state ($n_k = 0$ shown for reference), the critical-to-critical quench ($h_0 = h = 1$, $\gamma_0 = -1$ and $\gamma = 1$) and the noncritical-to-critical quench ($h_0 = 2$, $h = 1$ and $\gamma_0 = \gamma = 1$). At $k = 0$, there are different qualitative trends: in the former quench protocol, $n_{k=0} = 1$ while, in the latter quench protocol, $n_{k=0} = \frac{1}{2}$. Quenches starting from a (Middle) critical ($h_0 = 1, \gamma_0 = -1$) and (Right) noncritical ($h_0 = 2, \gamma_0 = 1$) initial states to the vicinity of the critical point ($h \approx 1, \gamma = 1$). Both cases exhibit a sharp discontinuity at the critical point $h = 1$: the occupation number $n_{k=0}$ becomes (Middle) 1/2, 1, 1/2 and (Right) 1, 1/2, 0 for $h < 1$, $h = 1$, and $h > 1$, respectively.

## S.II. METHODS FOR CALCULATING ENTANGLEMENT

To examine the critical correlations in the long-time stationary state of a quenched system, we investigate several measures of entanglement and information, including the von-Neumann entropy for a continuous region, mutual information as well as an upper bound on logarithmic negativity for adjacent regions, and the Rényi mutual information for disjoint regions.

### A. Von Neumann entropy

We first consider the von Neumann entropy $S_A = -\text{Tr}\rho_A \ln \rho_A$ of a connected region $A$ described by a (reduced) density matrix $\rho_A$. For a pure state, the von Neumann entropy gives the entanglement entropy of $A$ with the rest of the system. For the XY model, this quantity can be directly constructed from the correlation matrix $\Gamma_A$ [S4]. First, we remark that the eigenvalues of $\Gamma_A$ appear in pairs of $\pm \nu_i$. The von Nuemann entropy can be then computed as [S4]

$$S_A = -\sum_i \frac{1+\nu_i}{2} \ln \frac{1+\nu_i}{2} + \frac{1-\nu_i}{2} \ln \frac{1-\nu_i}{2}. \tag{S.35}$$

For a stationary state (including an excited state) characterized by the occupation numbers $\{n_k\}$, the Fisher-Hartwig conjecture [S6, S7] can be used to find an analytical expression for the von- Neumann entropy to the leading-order as [S8]

$$S_A = \frac{L_A}{2\pi} \int_0^{2\pi} dk \ \mathcal{H}(1 - 2\langle n_k \rangle) + \cdots, \tag{S.36}$$

where $L_A = |A|$, the function $\mathcal{H}(x) = -\frac{1+x}{2} \ln \frac{1+x}{2} - \frac{1-x}{2} \ln \frac{1-x}{2}$ defines the binary entropy, and the dots represent subleading correction terms (e.g., a logarithmic term and/or a constant). Equation (S.36) infers that any stationary state away from the ground state or the highest excited would unavoidably lead to an extensive entropy, proportional to $L_A$.

It is straightforward to calculate the mutual information for two adjacent subsystems, $A_1$ and $A_2$, from Eq. (S.35) as

$$I_{A_1:A_2} = S_{A_1} + S_{A_2} - S_{A_1 \cup A_2}. \tag{S.37}$$

On the other hand, computing the mutual information for disjoint regions is nontrivial. However, the Rényi mutual information for disjoint regions can be computed as we describe in Sec. S.II C.

### B. Upper bound on logarithmic negativity

Log-negativity ($\mathcal{E}$), an entanglement monotone, is defined as [S9, S10]

$$\mathcal{E}_{A_1:A_2} \equiv \ln ||\rho_A^{T_2}|| = \ln \text{Tr}|\rho_A^{T_2}|, \tag{S.38}$$

where $||\cdot||$ denotes the trace norm, and $\rho_A^{T_2}$ is the partial transpose of a reduced density matrix $\rho_A$, ($A = A_1 \cup A_2$) with respect to the subsystem $A_2$,

$$\rho_A^{T_2} = \langle e_i^{A_1} e_j^{A_2} | \rho_A^{T_2} | e_r^{A_1} e_s^{A_2} \rangle \equiv \langle e_i^{A_1} e_s^{A_2} | \rho_A | e_r^{A_1} e_j^{A_2} \rangle. \tag{S.39}$$

In the above equation, $|e_i^{A_1}\rangle$ and $|e_s^{A_2}\rangle$ define a basis in the Hilbert space $A_1$ and $A_2$, respectively. The log-negativity $\mathcal{E}_{A_1:A_2}$ is then calculated as the trace norm of $\rho_A^{T_2}$ which is equivalent to the sum of the absolute values of the eigenvalues of $\rho_A^{T_2}$ [S11].

Unfortunately, the partial transpose of a fermionic Gaussian state is, in general, not a Gaussian state [S11, S12]. The non-Gaussian nature of the partial transpose makes it difficult to efficiently calculate the log-negativity. However, by representing the partial transpose as a linear combination of Gaussian operators [S12], we can calculate a relatively tight upper bound $\hat{\mathcal{E}}_{A_1:A_2}$ ($\geq \mathcal{E}_{A_1:A_2}$) on log-negativity [S13]. $\hat{\mathcal{E}}_{A_1:A_2}$ exhibits the same critical scaling behavior as $\mathcal{E}_{A_1:A_2}$ at criticality [S14]. Specifically, for the XY model, the partial transpose of $\rho_A$ for two adjacent subsystems $A_1$ and $A_2$ can be written as a linear combination of two non-commuting Gaussian operators as [S11]

$$\rho_A^{T_2} = \frac{1-i}{2} \tilde{\rho}_A + \frac{1+i}{2} P_{A_2} \tilde{\rho}_A P_{A_2}$$
$$\equiv \frac{1-i}{2} O_+ + \frac{1+i}{2} O_-, \tag{S.40}$$

where $P_{A_2} = \prod_{j \in A_2}(ia_j^y a_j^x)$ represents a string of Majorana operators in the region $A_2$, and

$$\tilde{\rho}_A = \frac{1}{2^{L_A}} \sum_{\text{even, odd}} i^{\mu_2} \langle O_1 O_2 \rangle O_2^\dagger O_1^\dagger, \tag{S.41}$$

$$P_{A_2} \tilde{\rho}_A P_{A_2} = \tilde{\rho}_A^\dagger.$$

Here, $O_{1,2}$ denote a general product of Majorana fermions belonging to subsystem $A_{1,2}$, respectively, and $\mu_2$ is the number of Majorana fermions in $O_2$. The even/odd in the summation refer to an even/odd number of fermions in each block. The operators $O_+ \equiv \tilde{\rho}_A$ and $O_- \equiv P_{A_2} \tilde{\rho}_A P_{A_2}$ are each a Gaussian operator whose correlation matrices $\tilde{\Gamma}_1$ and $\tilde{\Gamma}_2$ can be evaluated as

$$\tilde{\Gamma}_k = \tilde{M}_2 \Gamma_k \tilde{M}_2, \qquad \tilde{M}_2 = \begin{bmatrix} \mathbf{1_L} & 0 \\ 0 & i\mathbf{1_L} \end{bmatrix}, \tag{S.42}$$

where $\mathbf{1_L}$ is an identity matrix of dimension $L = |A_1| = |A_2|$ and $\Gamma_1$ and $\Gamma_2$ are the correlation matrices corresponding to

$$\rho_A = \frac{1}{2^{L_A}} \sum_{\text{even, odd}} \langle O_1 O_2 \rangle O_2^\dagger O_1^\dagger,$$

$$P_{A_2} \rho_A P_{A_2} = \frac{1}{2^{L_A}} \left[ \sum_{\text{even}} \langle O_1 O_2 \rangle O_2^\dagger O_1^\dagger - \sum_{\text{odd}} \langle O_1 O_2 \rangle O_2^\dagger O_1^\dagger \right], \tag{S.43}$$

respectively. These correlation matrices can be explicitly evaluated as

$$\Gamma_1 = \Gamma^A, \quad \Gamma_2 = M_2 \Gamma_1 M_2, \tag{S.44}$$

where $\Gamma^A$ is the correlation matrix corresponding to the region $A$ [see Eq. (S.10)] and

$$M_2 = \begin{bmatrix} \mathbf{1_L} & 0 \\ 0 & -\mathbf{1_L} \end{bmatrix}. \tag{S.45}$$

While the log-negativity cannot be directly calculated for a non-Gaussian state, it follows from Eq. (S.40) that

$$||\rho_A^{T_2}|| \leq \left|\left|\frac{1-i}{2}O_+\right|\right| + \left|\left|\frac{1+i}{2}O_-\right|\right| = \sqrt{2}||O_+||, \tag{S.46}$$

where we have used the property $O_+^\dagger = O_-$; see Eq. (S.41). With this relation, an upper bound on log-negativity $\hat{\mathcal{E}}_{A_1:A_2}$ can be calculated as [S14, S15],

$$\hat{\mathcal{E}}_{A_1:A_2} = \ln \mathrm{Tr}(O_+ O_-)^{1/2} + \ln \sqrt{2}. \tag{S.47}$$

The trace norm of $O_+$ can be evaluated as

$$||O_+|| = \mathrm{Tr}(O_+ O_-)^{1/2} = \det\left[\left(\frac{\mathbb{1} + i\Gamma_x}{2}\right)^{1/2} + \left(\frac{\mathbb{1} + i\Gamma_x}{2}\right)^{1/2}\right] \det \frac{\mathbb{1} - \tilde{\Gamma}_1 \tilde{\Gamma}_2}{2}, \tag{S.48}$$

where $-i\Gamma_x = \mathbb{1} - (\mathbb{1} + i\tilde{\Gamma}_2)(\mathbb{1} - \tilde{\Gamma}_1 \tilde{\Gamma}_2)^{-1}(\mathbb{1} + i\tilde{\Gamma}_1)$.

### C. Disjoint Rényi mutual information

In this subsection, we show how the Rényi mutual information (RMI) can be computed for two disjoint regions. In general, the RMI between two systems $A_1$ and $A_2$ is given by

$$I^{(\alpha)}_{A_1:A_2} = R^{(\alpha)}_{A_1} + R^{(\alpha)}_{A_2} - R^{(\alpha)}_A, \tag{S.49}$$

where $R^{(\alpha)}_A = \frac{1}{1-\alpha} \mathrm{Tr} \rho_A^\alpha$ defines the Rényi entropy of the order $\alpha$, and $A = A_1 \cup A_2$. To calculate the RMI of two disjoint blocks (see the schematic picture in Fig. 3 of the main text), we first need to calculate the reduced density $\rho_A$. This density matrix can be constructed using the fact that only states with an even number of fermionic operators have non-vanishing expectation values [S16]:

$$\rho_A = \frac{1}{2^{L_A}} \left[ \sum_{\text{even}} \langle O_1 O_2 \rangle O_2^\dagger O_1^\dagger + \sum_{\text{odd}} \langle O_1 P O_2 \rangle O_2^\dagger P O_1^\dagger \right]. \tag{S.50}$$

Again, $L_A = |A|$ and $O_{1,2}$ represent the same operators defined in Eq. (S.41). The operator $P = \prod_{j \in B_1}(i a_j^y a_j^x)$ represents the string operator corresponding to the region $B_1$ defined as the region between $A_1$ and $A_2$. Now, since the spin variables on different sites commute, $[P, O_{1,2}] = 0$, we can write the density matrix as

$$\rho_A = \frac{1+P}{2} \rho_+ + \frac{1-P}{2} \rho_-, \tag{S.51}$$

where $\rho_\pm$ are fermionic reduced density matrices given by

$$\rho_\pm = \frac{1}{2^{L_A}} \left[ \sum_{\text{even}} \langle O_1 O_2 \rangle O_2^\dagger O_1^\dagger \pm \sum_{\text{odd}} \langle O_1 P O_2 \rangle O_2^\dagger O_1^\dagger \right] \tag{S.52}$$

$$= \rho_{\text{even}} \pm \rho_{\text{odd}}.$$

Note that $(1 \pm P)/2$ are orthogonal projectors, and commute with $\rho_\pm$. Next, we define

$$\rho_A^1 = \mathrm{Tr}_B \rho, \qquad \rho_A^P = \frac{\mathrm{Tr}_B(P\rho)}{\langle P \rangle}, \tag{S.53}$$

where $\rho$ is the fermionic density matrix of the total system, $B$ is the entire chain excluding the region $A$, and the trace is over fermionic observables in $B$; in the last equation, $\langle P \rangle$ is introduced to ensure normalization. Now, by using the string operator $P_1$ limited to the interval $A_1$, one can write

$$\rho_{\text{even}} = \frac{\rho_A^1 + P_1 \rho_A^1 P_1}{2}, \qquad \rho_{\text{odd}} = \frac{\rho_A^P - P_1 \rho_A^P P_1}{2}. \tag{S.54}$$

Using the above equations, the reduced density matrix $\rho_A$ can be written as a linear combination of four Gaussian operators with the corresponding correlation matrices given by

$$\Gamma_1 = \Gamma_{\rho_A^1}, \quad \Gamma_2 = \Gamma_{P_1 \rho_A^1 P_1}, \quad \Gamma_3 = \Gamma_{\rho_A^P}, \quad \Gamma_4 = \Gamma_{P_1 \rho_A^P P_1}. \tag{S.55}$$

These correlation matrices can be written explicitly as [S16]

$$\Gamma_1 = \Gamma_{AA}, \quad \Gamma_2 = M_2 \Gamma_1 M_2, \quad \Gamma_3 = \Gamma_{AA} - \Gamma_{AB_1} \Gamma_{B_1 B_1}^{-1} \Gamma_{B_1 A}, \quad \Gamma_4 = M_2 \Gamma_3 M_2, \tag{S.56}$$

where the matrix $M_2$ is given in Eq. (S.45) and the first (second) symbol in the subscript of the correlation matrices indicate the region over which the row (column) index runs. These correlation matrices can be calculated using Eq. (S.10) with the row and column indices running over the respective regions indicated by the subscript.

Using these expressions, we can construct $\text{Tr}(\rho^\alpha)$ for integer $\alpha > 0$. However, computing the mutual information (by setting $\alpha \to 1$) for disjoint regions is nontrivial, and we shall not pursue it here. Instead, we consider $\alpha = 2$; we expect the same qualitative behavior for different values of $\alpha$ as well. We have we have [S16]

$$R^{(2)}_{A_1} + R^{(2)}_{A_2} = -\ln(\{\bar{\Gamma}^2\}), \qquad \bar{\Gamma} = \frac{\Gamma_1 + \Gamma_2}{2}, \tag{S.57a}$$

$$R^{(2)}_A = -\ln\left[\frac{1}{2}(\{\Gamma_1^2\} + \{\Gamma_1, \Gamma_2\} + \det(\Gamma_{B_1})(\{\Gamma_3^2\} - \{\Gamma_3, \Gamma_4\}))\right], \tag{S.57b}$$

where $\Gamma_{B_1}$ is the correlation matrix corresponding to the region $B_1$ and the bracket is defined as

$$\{\Gamma, \Gamma'\} = \sqrt{\det\left(\frac{1 + \Gamma\Gamma'}{2}\right)}, \tag{S.58}$$

with the convention that $\{\Gamma^2\} \equiv \{\Gamma, \Gamma\}$. Putting these pieces together in Eq. (S.49), we can construct the disjoint RMI, $I^{(2)}_{A_1:A_2}$.

### 1. Critical-to-critical quench: Asymptotic behavior of the disjoint RMI

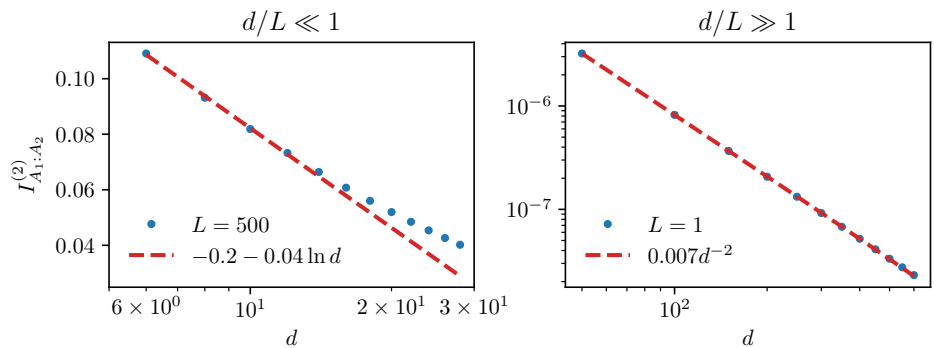

FIG. S.3. Disjoint Rényi mutual information (RMI) $I^{(2)}_{A_1:A_2}$ for a critical-to-critical quench ($h_0 = h = 1$, $-\gamma_0 = \gamma = 1$) plotted for (Left) $d/L \ll 1$ and (Right) $d/L \gg 1$. At short separations, the RMI scales logarithmically with $d/L$ while at large separations, it decays algebraically proportional to $d^{-2}$.

For a critical-to-critical quench, the disjoint RMI, $I^{(2)}_{A_1:A_2}$ for subsystems $A_1$ and $A_2$ each of size $L$ and separated by a distance $d$, is shown in Fig. 3 of the main text. Here, in Fig. S.3, we report the asymptotic scaling in two opposite

regimes where the distance $d$ is much smaller or larger compared to the regions' size $L$. Our numerical results are consistent with the scaling

$$I^{(2)}_{A_1:A_2} \propto \begin{cases} \ln(L/d), & \text{for } d/L \ll 1, \\ (L/d)^2, & \text{for } d/L \gg 1. \end{cases} \quad (S.59)$$

The constants of proportionality and the additive constant subleading to the logarithmic scaling are reported in Fig. S.3. We mention in passing that the mutual information itself falls off as $(L/d)^4$ at large distances, displaying a contrast with the RMI as pointed out in Ref. [S17]; see also Sec. S.IV. The long-distance scaling is due to the same power-law decay of the connected transverse correlation functions; longitudinal correlations decay exponentially in this case. The logarithmic scaling at short distances also arises in the critical ground state [S18]. In both cases, the coefficient of the logarithmic term is expected to be $1/8$ for the second RMI. However, smaller values of $d/L$ and a more careful finite-size scaling analysis are necessary to capture the asymptotic behavior [S19].

### 2. Critical-to-noncritical quench: Lack of scaling invariance of the RMI

We have shown in Fig. 3 of the main text that the disjoint RMI exhibits a scaling collapse for the critical-to-critical quench, while it is not scale-invariant for the noncritical-to-critical quench. Here, we show that a critical-to-noncritical quench too shows no such scale invariance. Figure S.4 displays $I^{(2)}_{A_1:A_2}$ as a function of $d/L$ and for different system sizes. Indeed, the data does not show a scaling collapse. We note that the RMI in Fig. S.4 appears to decay exponentially although the connected transverse correlations decay algebraically as $\rho^{zz}_l \propto l^{-4}$; see Eq. (S.28). While this power law decays rather slowly compared to an exponential decay of longitudinal correlations, it is numerically smaller than the latter in the regime $d/L$ shown in the figure.

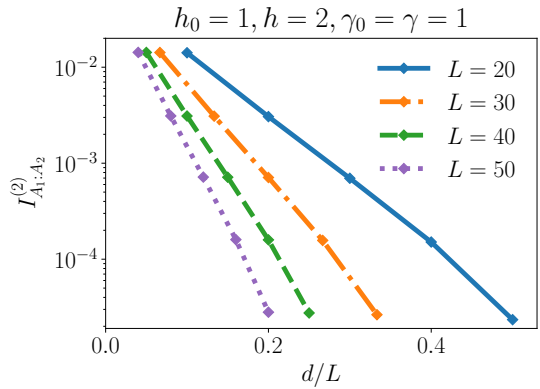

FIG. S.4. Disjoint Rényi mutual information $I^{(2)}_{A_1:A_2}$ for a critical-to-noncritical quench ($h_0 = 1, h = 2, \gamma_0 = \gamma = 1$). The data does not show a scaling invariance, hence the absence of any critical behavior.

## S.III. MUTUAL INFORMATION AND LOG-NEGATIVITY: SCALING WITH $|h-1|$ AND $L$

The mutual information and log-negativity display a strong dependence on the system size upon a quench to the critical point for both quench protocols; see Fig. 1(e-h) of the main text. These quantities are increasing for larger subsystem sizes $L$. This behavior is expected for the mutual information as a consequence of the strong subadditivity property as long as the $L$ is smaller than half the total system size (trivially satisfied in the thermodynamic limit) [S20]. In this section, we focus on the strong subsystem-size ($L$) dependence of both the mutual information and log-negativity near the (post-quench) critical point.

We start with the first quench protocol where the initial state is noncritical while the final state is in the vicinity of the critical point, $0 < |h - h_c| \ll 1$. To single out the critical contribution, we consider the limit $L \to \infty$ where both the mutual information $I_{A_1:A_2} \equiv I(L, h)$ and the (upper bound on) log-negativity $\hat{\mathcal{E}}_{A_1:A_2} \equiv \hat{\mathcal{E}}(L, h)$ approach a finite, $L$-independent value, $I(\infty, h)$ and $\hat{\mathcal{E}}(\infty, h)$, respectively. We then analyze the finite-size scaling for large, but

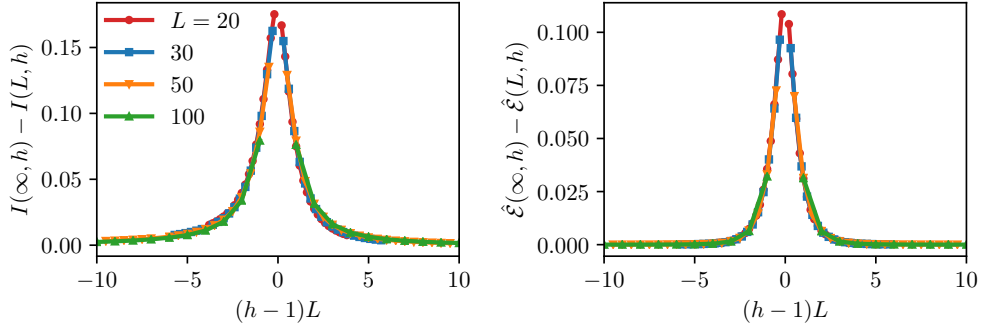

FIG. S.5. Finite-size scaling of (Left) the mutual information and (Right) the upper bound on log-negativity for a quench from a noncritical initial Hamiltonian to the vicinity of a critical point ($h_0 = 2, h \approx 1, \gamma_0 = \gamma = 1$). Here, the value of the mutual information and log-negativity in the thermodynamic limit, $I(\infty, h)$ and $\hat{\mathcal{E}}(\infty, h)$, are approximated by their respective values for $L = 500$. We have not plotted the data for $h = 1$ as both $I(L, h)$ and $\hat{\mathcal{E}}(L, h)$ are independent of the subsystem size $L$ exactly at $h = h_c$. The data in both panels are shown for the subsystem sizes specified in the left panel.

finite $L$. In Fig. S.5, we show that a scaling collapse is possible by introducing

$$I(L, h) = I(\infty, h) - \mathcal{F}\big[(h-1)L\big], \qquad (\text{S.60})$$
$$\hat{\mathcal{E}}(L, h) = \hat{\mathcal{E}}(\infty, h) - \mathcal{G}\big[(h-1)L\big], \qquad (\text{S.61})$$

with $\mathcal{F}(x)$ and $\mathcal{G}(x)$ the corresponding scaling functions which are always finite and approach 0 as $x \to \infty$. Exactly at the critical point, $h = 1$, the mutual information (log-negativity) takes the same value independent of the subsystem size $L$, and thus the scaling functions $\mathcal{F}(x)$ and $\mathcal{G}(x)$ become 0 for $x = 0$ as well; however, there is a discontinuity as these functions are nonzero in the limit $x \to 0$. As pointed out in the main text and detailed here, the scaling behavior, the sharp dip, and the discontinuity at $h = 1$ reflect the behavior of the occupation number; c.f. Eq. (S.34).

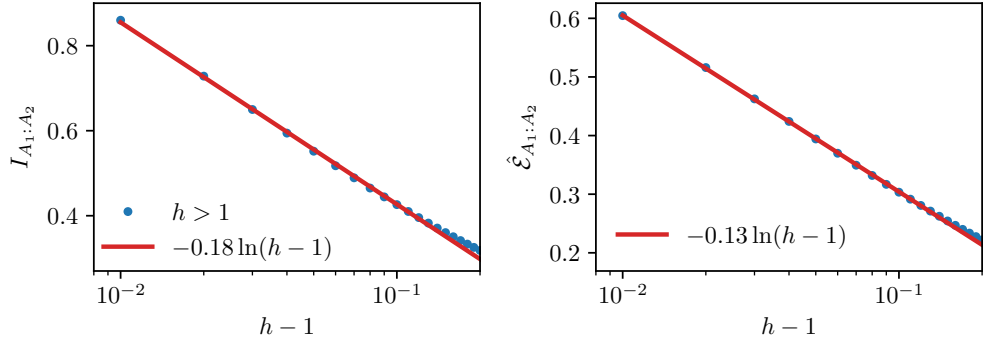

FIG. S.6. Scaling of (Left) the mutual information and (Right) log-negativity with the distance from the critical point $h - 1$ for a quench from a critical initial Hamiltonian to the vicinity of another critical point ($h_0 = 1, h \approx 1, -\gamma_0 = \gamma = 1$); we have chosen subsystem size $L = 450$. The scaling is consistent with $I_{A_1:A_2} \sim -\frac{1}{6}\ln(h-1)$ and $\hat{\mathcal{E}}_{A_1:A_2} \sim -\frac{1}{8}\ln(h-1)$.

Next, we consider the second protocol where the initial state is critical and a quench is considered to the vicinity of another critical point. In this case, the mutual information and log-negativity exhibit a strong dependence on the subsystem size, and furthermore they diverge at the (post-quench) critical point. Here, we focus on the dependence of these quantities on the distance away from the critical point, $h - 1$. Figure S.6 shows that, for $h > 1$, these quantities diverge logarithmically as $-b \ln(h-1)$ where the coefficients are consistent with $b = \frac{1}{6}$ for the mutual information and $b = \frac{1}{8}$ for the log-negativity. These scaling behaviors and the crossover between them follow from that of the occupation number; c.f. Eq. (S.30).

## S.IV. CORRELATIONS $\sim 1/l^\alpha$ WITH $\alpha > 1$: AREA LAW FOR MUTUAL INFORMATION

In the main text, we have stated that an algebraic decay of the connected transverse correlations such as $\langle \sigma_j^z \sigma_{j+l}^z \rangle_c \sim 1/l^2$ cannot lead to a logarithmic violation of the area law assuming that longitudinal correlations decay exponentially and that higher-order correlations can be neglected as well. Here, we present an argument for the area law under these assumptions and a for a more general case where correlations decay as $1/|l|^\alpha$ for $\alpha > 1$. For simplicity, however, we consider only "classical" correlations where only $\sigma^z$ correlators decay algebraically while correlators involving $\sigma^{x,y}$ are simply set to zero. As long as the latter decay exponentially, our assumption is a good approximation at long distances; a logarithmic scaling of the mutual information, if any, diverges for large subsystem sizes and thus is most sensitive to long-range correlations. We thus introduce a "classical" fictitious Hamiltonian $H = -\sum_{i \neq j} J_{ij} \sigma_i^z \sigma_j^z$ at the inverse temperature $\beta$. Assuming that $J_{ij} \sim J/|i-j|^\alpha$, the correlation function at sufficiently high temperatures ($\beta J \ll 1$) decay as

$$\langle \sigma_i^z \sigma_j^z \rangle \sim \frac{\beta J}{|i-j|^\alpha}, \qquad \text{when } \beta J \ll 1. \tag{S.62}$$

Here, we have assumed that $\langle \sigma_i^z \rangle = 0$; we will shortly discuss the extension to the more general case. Notice that higher-order correlations are smaller by higher orders of $\beta J$. We also remark that we have limited ourselves to $\alpha > 1$ since the fictitious Hamiltonian $H$ becomes non-extensive for any $\alpha \leq 1$.

Next, we consider two regions $A$ and $B$ on the chain; for example, $A$ can be a contiguous set of spins in a region of size $L$ and $B$ its complement on an infinite chain. It is convenient to write the Hamiltonian as $H = H_A + H_B + H_{\text{int}}$ where $H_{A(B)}$ only involves spins that are strictly in $A$ ($B$) and $H_{\text{int}}$ involves terms that cross the boundary between $A$ and $B$. We now use a result from Ref. [S20]: for a thermal state, the mutual information $I(A, B)$ is bounded by

$$I_{A:B} \leq \beta \text{Tr}\left[ H_{\text{int}}(\rho_A \otimes \rho_B - \rho_{AB}) \right], \tag{S.63}$$

where $\rho_R$ is the (reduced) density matrix of the region $R$. This inequality simply follows from the fact that the free energy is minimized for a thermal state. Inserting the Hamiltonian $H_{\text{int}}$ in the above equation, we find

$$I_{A:B} \leq \beta \sum_{i \in A, j \in B} J_{ij} \langle \sigma_i^z \sigma_j^z \rangle_c. \tag{S.64}$$

Notice that the connected correlation function appears in this equation, although at high temperatures and due to the symmetry of the fictitious Hamiltonian, we have $\langle \sigma_i^z \rangle = 0$. We then obtain

$$I_{A:B} \leq (\beta J)^2 \sum_{i \in A, j \in B} \frac{1}{|i-j|^{2\alpha}} \leq \zeta(2\alpha - 1) (\beta J)^2, \qquad \text{when } \beta J \ll 1, \tag{S.65}$$

where $\zeta$ is the Riemann zeta function. Given that $\alpha > 1$, the mutual information is bounded by a constant, independent of the size of $A$ or $B$, hence the area law. Our argument is straightforwardly extended to a scenario where there is a magnetic field $h \sum_j \sigma_j^z$. In this case, $H_{\text{int}}$ remains the same and the connected correlation appearing in Eq. (S.64) decay in the same fashion as before.

The perturbative nature of this argument as a high-temperature expansion requires the prefactor in Eq. (S.62) to be small. From Eq. (S.21) for the critical-to-critical quench, we have $\rho_l^{zz} \sim 1/\pi^2 l^2$ (taking $\gamma = 1$), therefore $(\beta J)^2 \equiv 1/\pi^2 \ll 1$ is a good approximation. We remark that an area law at finite temperature can be generally proved for any $\alpha > 2$ without any assumptions [S21].

---


\end{document}